\documentclass[twocolumn,prd,noshowpacs,nofootinbib,amsmath,amssymb,superscriptaddress]{revtex4}
\usepackage{graphicx}
\usepackage{amssymb}
\usepackage{amsmath}
\usepackage{amsfonts}
\usepackage{latexsym}
\usepackage{hyperref}
\usepackage[all]{xy}
\usepackage{slashed}
\newcommand{\be}{\begin{eqnarray}}
\newcommand{\ee}{\end{eqnarray}}

\newcommand{\GeV}{~\mathrm{GeV}}
\newcommand{\TeV}{~\mathrm{TeV}}

\newcommand{\thetaW}{\theta_{_W}}
\newcommand{\ZZ}{{\rm Z}^0}

\newcommand{\MQ}{M_{\rm mCP}}
\newcommand{\degree}{^{o}}
\newcommand{\gsim}{\lower.7ex\hbox{$\;\stackrel{\textstyle>}{\sim}\;$}}
\newcommand{\lsim}{\lower.7ex\hbox{$\;\stackrel{\textstyle<}{\sim}\;$}}

\begin{document}

\title{Looking for milli-charged particles with a new experiment at the LHC}
\author{Andrew Haas}
\affiliation{Department of Physics, New York University, New York, NY, USA}
\author{Christopher S. Hill}
\affiliation{Department of Physics, The Ohio State University, Columbus, OH, USA}
\author{Eder Izaguirre}
\affiliation{Perimeter Institute for Theoretical Physics, Waterloo, Ontario, Canada}
\author{Itay Yavin}
\affiliation{Perimeter Institute for Theoretical Physics, Waterloo, Ontario, Canada}
\affiliation{Department of Physics, McMaster University,  
Hamilton, ON, Canada}

\begin{abstract}
We propose a new experiment at the Large Hadron Collider (LHC) that offers a powerful and model-independent probe for milli-charged particles. This experiment could be sensitive to charges in the range $10^{-3}e - 10^{-1}e$ for masses in the range $0.1 - 100 \GeV$, which is the least constrained part of the parameter space for milli-charged particles. This is a new window of opportunity for exploring physics beyond the Standard Model at the LHC. The key new ingredients of the proposal are the identification of an optimal location for the detector and a telescopic/coincidence design that greatly reduces the background. 
\end{abstract}

\maketitle
With no clear evidence for the existence of grand unification or magnetic monopoles, the quantization of charge remains a mystery~\cite{Langacker:1980kd}. The search for a non-quantized charged particle, commonly known as a {\it milli-charged particle} (mCP),  progressed through the efforts of many direct experiments and indirect observations over many years~\cite{Davidson:1991si,Davidson:2000hf,Essig:2013lka}. The parameter space spanned by the mass and charge of the mCPs is strongly constrained by direct searches from accelerator experiments \cite{Prinz:1998ua, Davidson:2000hf, Badertscher:2006fm, CMS:2012xi} and indirect observations from astrophysical systems \cite{Davidson:1991si, Mohapatra:1990vq, Davidson:1993sj, Davidson:2000hf}, the cosmic microwave background \cite{Dubovsky:2003yn, Dolgov:2013una}, big-bang nucleosynthesis \cite{Vogel:2013raa}, and universe over-closure bounds \cite{Davidson:1991si}. While direct laboratory searches robustly constrain the parameter space of mCPs, indirect observations can be evaded by adding extra degrees of freedom. In particular, the parameter space for mCPs with masses $\MQ$ $0.1 \lsim\MQ\lsim100\GeV$ is largely unexplored by direct searches.

In this {\it Letter} we propose a new search to be conducted at the LHC with a dedicated detector targeting this unexplored part of parameter space, namely mCP masses $0.1 \lsim \MQ \lsim 100 \GeV$, for charges $Q$ at the $10^{-3}e - 10^{-1}e$ level. The experimental apparatus would be one or more roughly 1 m$^3$ scintillator detector layers positioned near one of the high-luminosity interaction points of the LHC, such as ATLAS or CMS. 
The experimental signature would consist of a few photo-electrons (PE) arising from the small ionization produced by the mCPs that travel unimpeded through material after escaping the ATLAS or CMS detectors. Moreover, the experiment proposed 
is a model-independent probe of mCPs, since it relies {\it only} on the production and detection of mCPs through their QED interactions.

We base the estimates for the potential reach of our proposed experiment on a particular theoretical framework, which we now briefly describe. While it is possible to simply add mCP particles to the Standard Model (SM), this is both unappealing from a theoretical point of view as well as strongly constrained by early universe over-production of these particles (see~\cite{Davidson:1991si, Davidson:2000hf, Langacker:2011db,Vogel:2013raa} and references therein). A more appealing possibility is the existence of an extra abelian gauge field that couples to a massive Dirac fermion (``dark QED") and that mixes with hypercharge through the kinetic term~\cite{Holdom:1985ag},
\be
\label{eqn:def_lag}
\nonumber
\mathcal{L} &=& \mathcal{L}_{\rm SM} -\frac{1}{4}A'_{\mu\nu}A^{\prime\mu\nu} + i\bar{\psi}\left( \slashed{\partial}+ie' \slashed{A}' +i \MQ\right)\psi \\ &~& -\frac{\kappa}{2} A'_{\mu\nu}B^{\mu\nu} .\
\ee
Here $\psi$ is a Dirac fermion of mass $\MQ$ that is charged under the new $U(1)$ field $A'_{\mu}$ with charge $e'$, and the field-strength is defined as $A'_{\mu\nu} = \partial_\mu A'_\nu - \partial_\nu A'_\mu$. The last term in Eq.~(\ref{eqn:def_lag}) is a kinetic mixing term between the field strength of the new gauge boson and that of hypercharge. Such a term is expected in grand unified theories and more generally whenever there exists massive fields that are charged under both hypercharge and the new gauge boson, even when these heavy fields are not accessible at low energies. 

Eliminating the mixing term by redefining the new gauge boson as, $A'_\mu \rightarrow A'_\mu + \kappa B_\mu$  results in a coupling of the charged matter field $\psi$ to hypercharge (as well as an immaterial redefinition the hypercharge coupling), 
\be
\nonumber
\mathcal{L} &=& \mathcal{L}_{\rm SM}  -\frac{1}{4}A'_{\mu\nu}A^{\prime\mu\nu} \\ &+& i\bar{\psi}\left( \slashed{\partial}+ie' \slashed{A}' - i \kappa e' \slashed{B} + i \MQ \right)\psi .\
\label{eqn:mCP_lag}
\ee
The new matter field $\psi$ therefore acts as a field charged under hypercharge with a charge $\kappa e'$, a milli-charge~\cite{Holdom:1985ag}. The mCP $\psi$ couples to the photon and $\ZZ$ boson with a charge $\kappa e'\cos\thetaW $ and $-\kappa e'\sin\thetaW $, respectively. The fractional charge in units of the electric charge is therefore $\epsilon \equiv \kappa e' \cos\thetaW/e$. 
 
\begin{figure}[t]
\centering
\includegraphics[width=0.48 \textwidth ]{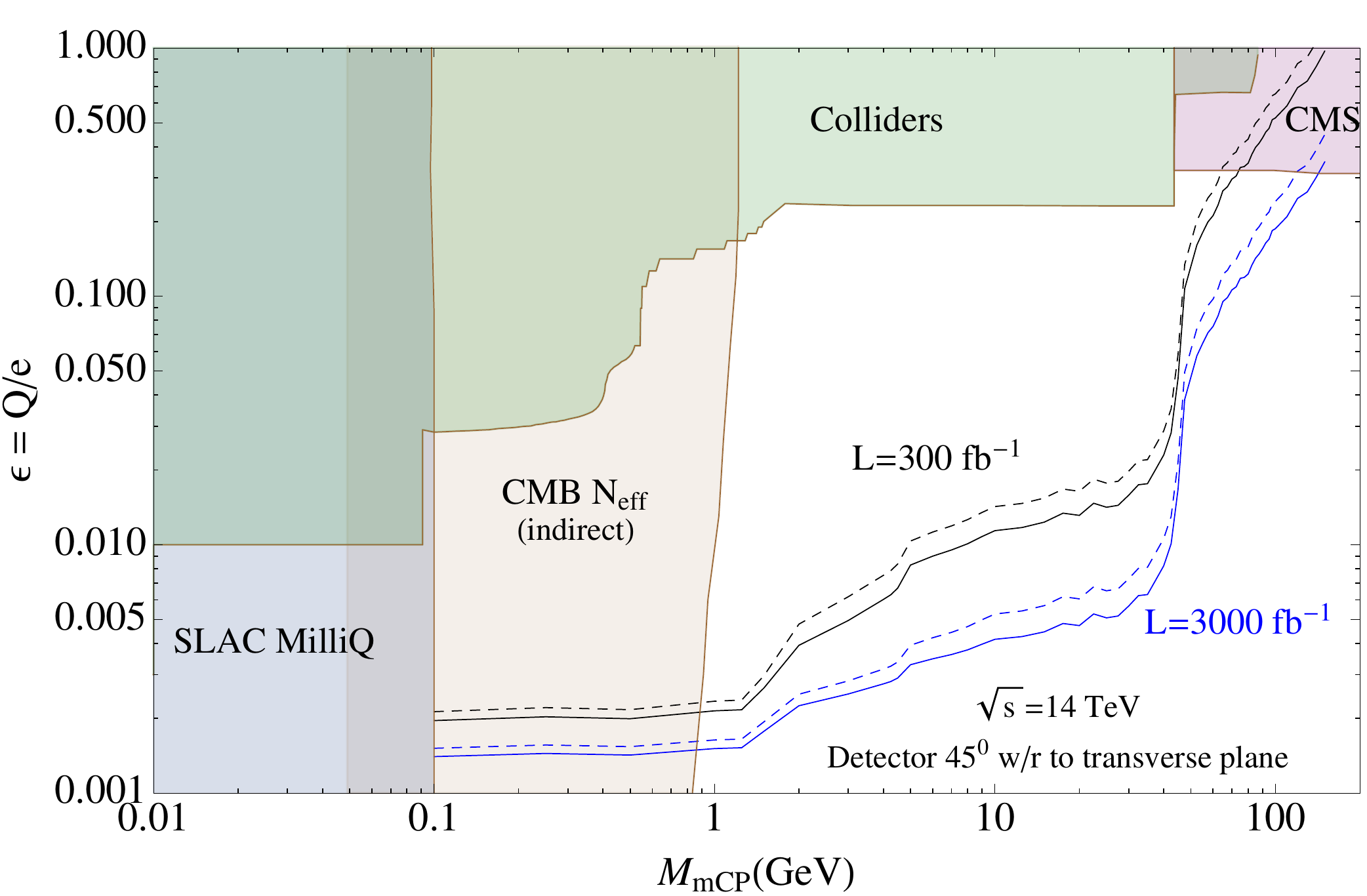}
\caption{The direct and indirect bounds on mCPs for models with a massless dark photon as well as the projected reach of the current proposal. Existing bounds are taken from Fig. 1 of ref.~\cite{Vogel:2013raa} and are explained in more details in the text. For $\MQ < 0.1 \GeV$, direct constraints were placed by the dedicated search at SLAC (blue) \cite{Prinz:1998ua} and by the search for invisible decays of ortho-positronium \cite{Badertscher:2006fm}.  For $\MQ > 0.1 \GeV$ --- the parameter space relevant to our proposal --- the strongest direct bounds arise from colliders (shaded green), which consist of beam-dump experiments, the constraint from the invisible width of the $Z$, as well as direct searches for fractionally charged particles at LEP \cite{Davidson:2000hf}. A recent direct search by CMS is shown in light purple \cite{CMS:2012xi,Jaeckel:2012yz}. The black line shows the expected 95\% C.L. exclusion (solid) and $3\sigma$ sensitivity (dashed) of the proposed experiment at $\sqrt{s}=14\TeV$, assuming $300/\rm fb$ of integrated luminosity. In blue we show the corresponding lines achievable in the high-luminosity LHC run with $3000/\rm fb$. }
\label{fig:mQBounds}
\end{figure}

In Fig.~\ref{fig:mQBounds} we show the general direct constraints on mCPs in this scenario~\cite{Vogel:2013raa}, together with the projected sensitivity associated with the current proposal. Some indirect constraints are not presented since they cannot be employed without additional assumptions (such as knowledge of the flux of mCPs on Earth), but we refer the reader to~\cite{Perl:2009zz} for an excellent review of all the different indirect searches. For the range of couplings relevant for this proposal, the mCP particles are abundantly produced in the early Universe. However, as the Universe cools, the number density of mCP is exponentially depleted through pair annihilation in the dark sector. This process can be sufficiently efficient to avoid the different constraints on relic abundance of the mCP~\cite{Davidson:1991si, Dubovsky:2003yn, Dolgov:2013una,Vogel:2013raa}. For example, with $e'=0.3$ and $\MQ=1\GeV$, the relic density is $10^{-5}$ that of dark matter. Even smaller relic densities are expected for a larger dark charge or if more annihilation channels are open. Another indirect constraint comes from the the number of relativistic degrees of freedom bound, known as $N_{\rm eff}$. However, since the dark photon decouples together with the mCP it will be much colder than the rest of the SM after entropy injection at later times. Since the contribution to $N_{\rm eff}$ scales like the fourth power of the temperature, this contribution is negligible when decoupling happens above a GeV or so~\cite{Jaeckel:2010ni, Brust:2013ova,Vogel:2013raa}.

Laboratory experiments have placed the strongest direct limits on mCPs for $10^{-5} \lsim \epsilon \lsim 10^{-1}$ over the range $\MQ < 300\GeV$. These constraints are the result of a dedicated search for mCPs at SLAC \cite{Prinz:1998ua}; collider experiments consisting of dedicated searches at beam-dump experiments,  free-quark searches, trident process searches, constraints from the invisible width of the $Z$ as well as direct searches for fractionally-charged particles at LEP \cite{Davidson:2000hf}; and decays of ortho-positronium \cite{Badertscher:2006fm}. A recent analysis looking for low ionizing particles in CMS excluded particles with charge $\pm e/3$ for $\MQ < 140\GeV$ and particles with charge $\pm2e/3$ for $\MQ < 310\GeV$~\cite{CMS:2012xi}.

Two other possible new probes of mCPs with $\MQ \gsim \GeV$ are cosmic rays and B-factories. The flux of mCPs with a mass and a charge in the range of interest to this proposal originating from cosmic rays is negligible for detection at either surface or underground experiments. B-factories offer two  classes of searches with potential sensitivity. First, searches for tagged mesons decaying invisibly (for example $\Upsilon(1s)\rightarrow\text{invisible}$) could be sensitive to the process $b\bar b \rightarrow \psi \bar\psi$, where the $\psi$'s are registered as missing energy. However, the sensitivity of these analyses is relatively weak \cite{Aubert:2009ae}. In particular, we find $Q<0.23e$ for $\MQ < m_{\Upsilon(1s)}/2$.  Another possibility is the continuum process $e^+e^-\rightarrow \gamma\psi\bar\psi$. We recast the results from the BaBar search for the untagged decay $\Upsilon(3S)\rightarrow\gamma A^0$, with $A^0$ an invisibly decaying scalar \cite{Aubert:2008as}. The analysis consists of a bump-search in the observable $m_X^2\equiv m_{\Upsilon(3s)}^2-2E_\gamma^*m_{\Upsilon(3s)}$, where $E_\gamma^*$ is the photon energy in the centre-of-mass frame. Because the analysis is a bump search the sensitivity achieved by simply recasting the results to the milli-charge continuum process was found to be sub-optimal: at a mass of $\MQ = 0.1\GeV$ the mCP's charge is only constrained to be less than $0.1e$ and the bound deteriorates quickly to as high as $0.5e$ at $\MQ = 0.5\GeV$.

\begin{figure}[t]
\centering
\includegraphics[width=0.45\textwidth]{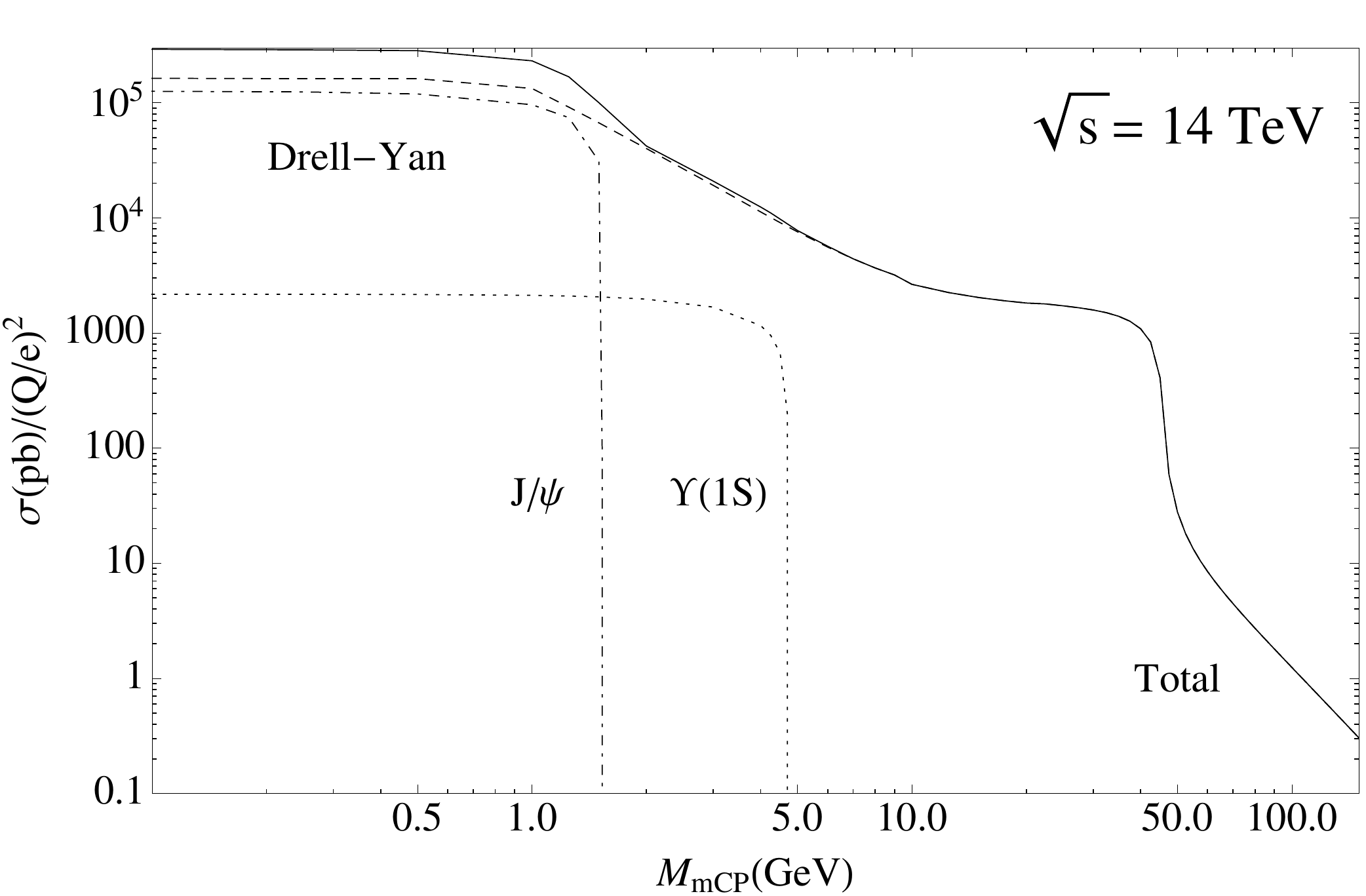}
\caption{Cross-section at the LHC for the process $pp \rightarrow \psi\bar\psi+X$ as a function of the 
mCP mass. As described in the text, the combined cross-section (solid) is obtained through Drell-Yan (dashed), $\Upsilon(1s)\rightarrow \psi\bar\psi+X$ (dotted), and $J/\psi \rightarrow \psi\bar\psi+X$ (dot-dashed). The enhancement for $\MQ < m_{Z^0}/2$ is due to the $Z^0$-mediated contribution.}
\label{fig:prodxs}
\end{figure}

We now turn our discussion to the LHC, which offers a unique target of opportunity for model-independent searches for mCPs in the $0.1 - 100 \GeV$ range. First we address the production mechanism for mCPs, followed by a discussion of the experimental setup for detection.
%
%
The main production channel at hadron colliders for mCPs in the GeV mass range is through the Drell-Yan process. We show the production cross section for mCPs in ``dark QED'' in Fig.~\ref{fig:prodxs}. With the expected instantaneous luminosities of Run2 of the LHC (up to $\mathcal{L}\approx 2\times10^{34}$~cm$^{-2}$s$^{-1}$), one is faced with tens to thousands of mCPs produced in the mass and charge range considered in this proposal. We simulated the mCPs Drell-Yan signal in {\sc Madgraph5}~\cite{Alwall:2011uj}, where they can be produced through an s-channel $\gamma$ or $Z^0$, after implementing our own model where we added a new fermion to the SM charged under hypercharge only, as in Eq.~(\ref{eqn:mCP_lag}). Fig.~\ref{fig:prodxs} shows the production cross section for mCP pairs with combined invariant mass greater than $2 \GeV$ (the lowest momentum transfer where PDFs have been studied~\cite{Ball:2012cx}). Moreover, when quoting the bounds we include a k-factor of 1.3 from the NNLO normalization, which we obtain from \textsc{Vrap}~\cite{vrap}. We also included two other significant mCP production channels through the decay of quarkonia:  $\Upsilon(1s)\rightarrow \psi\bar\psi$ and $J/\psi \rightarrow \psi\bar\psi$. The $\Upsilon(1s)\rightarrow \psi\bar\psi+X$ cross section includes $y_{\Upsilon(1s)} < 2.5$ and $p_{T,\Upsilon(1s)} > 2 \GeV$. The $J/\psi \rightarrow \psi\bar\psi+X$ is shown after requiring $y_{J/\psi} < 2.5$ and $p_{T, J/\psi} > 6.5 \GeV$. We modeled these with \textsc{MadOnia} \cite{Artoisenet:2007qm}, and rescaled the branching ratio by $\epsilon^2$ and the appropriate phase-space factor associated with the mass of the mCP.

We now move on to describe the experimental considerations that go into this proposal. mCPs with mass above 100 MeV will lose energy primarily through ionization and excitation, as described by the Bethe-Bloch formula, with a factor $Q^2$ compared to a standard minimum-ionizing charged particle. Fractionally charged particles leave very little energy deposition in any standard particle detector at the LHC and do not register signals above threshold for roughly Q$< 0.3e$~\cite{CMS:2012xi}.  Pair-produced mCPs would thus constitute missing energy, but can be searched for using a mono-X search that exploits initial state radiation to produce a triggerable energy deposition (or equivalently missing transverse momentum). However, for electroweak production of new invisible states, such as the mCPs studied here, the penalty on the cross-section associated with initial state radiation precludes CMS/ATLAS sensitivity to such particles, even with the very large datasets envisaged for the HL-LHC.  For instance, the production cross-section of a mCP-pair with $\MQ=10\GeV$ in association with a jet of $p_T > 250 \GeV$  at $\sqrt{s}=14\TeV$ is $\approx \epsilon^2\times$0.5 pb. Assuming the same $p_T$ cut on the jet, the cross section for ${\bar \nu}\nu+1j$ through the $Z^0$ boson is $\sim 3.5$ pb at 14 TeV. This background alone would require an integrated luminosity of greater than 300~fb$^{-1}$ to yield a ${\rm S/\sqrt{B}}\sim $ a few for $Q=0.1e$. Additionally, the small ${\rm S/B} \sim 10^{-4}$ means that more data will not necessarily improve the sensitivity as the analyses become systematics-dominated. Thus, to detect mCPs at the LHC, an alternative experimental strategy is likely needed. 

We thus propose a dedicated experiment to search for mCPs in which one or more small (1 $\rm m^3$) scintillator layers are deployed sufficiently near one of the high-luminosity LHC interactions points, i.e.\ ATLAS/CMS, such that a detectable fraction of the flux of mCPs produced in the $pp$ collisions provided by the LHC would be intercepted by the experimental apparatus. The rapidity distribution in the lab frame extends to high values even for $\MQ \sim 50\GeV$. The fraction of events with at least one mCP in the rapidity range $|\eta|<1$ for $\MQ = 1,5,10$ and $50\GeV$ is $12\%, 19\%, 23\%$, and $27\%$, respectively. Given the expected rapidity distribution, better coverage would be obtained by placing the detector as much in the forward region as possible. At the same time, since the flux generally drops with the square of the distance one would like to place the detector as close as possible to the pp interaction point of either ATLAS or CMS. 

A mCP detector must be shielded from the ionizing radiation produced by SM particles emanating from the proton beams and their interactions. Because a mCP detector will necessarily be designed to be sensitive to extremely small ionization energy depositions, any $Q=1e$ particle entering the detector will flood it with photons for up to several $\mu$s, during which no mCP signal could be seen. The large flux of such particles within the ATLAS/CMS experimental caverns, therefore, rules out placement of a mCP detector there.  Moreover, even if it were a suitable environment, there is no sufficient space available in the already crowded forward regions of these experiments.
One possibility would be to locate detectors on the surface, roughly 100 m above the interaction point. Another more advantageous possible location would be to use the counting room adjacent to the experimental cavern, typically located underground and about 20 m from the interaction point. For instance, USA15, near ATLAS, is a large cavern housing computing and trigger electronics shielded by a 2 m thick concrete wall that reduces radiation to less than a few $\mu$Sv/h \cite{Dawson:2004pta}, only about 10 times the normal environmental background rate. A similar environment exists in USC55, near CMS.  An advantage of the low-radiation requirement of the counting rooms is that access to these experimental areas is possible during running beam conditions. The acceptance of a 1 $\rm m^2$ detector at such a location 20 m away would be about 0.01\% for $\MQ\sim$ few GeV.

A $Q=1e$ minimum-ionizing charged particle leaves roughly 2~MeV/cm in a material of density 1 g/cm$^3$~\cite{Beringer:1900zz}. For plastic scintillator, such energy deposition results in about $10^4$ photons per MeV, meaning $2\times10^6$ photons would be liberated in a 1~m long scintillator. For a particle with electric charge $Q<1e$, the energy deposition is reduced by the factor of $Q^2$ mentioned above, resulting in just a few photons liberated on average in the same 1~m long scintillator. Allowing for an estimated detection efficiency of about 10\%, we can therefore expect an average of one PE via an attached phototube for each mCP with $Q=2\times10^{-3}e$ that traverses a standard length of 140~cm plastic scintillator~\cite{Prinz:1998ua}. The signal to search for then is one or more PEs.

Similar to Ref.~\cite{Prinz:1998ua}, 140~$\times$~10~$\times$~5~cm plastic scintillator bars could be used, with an associated phototube and readout for each bar. 200 such bars would be needed to cover 1 m$^2$ of area perpendicular to the beam-line, with each mCP passing through the 140~cm length of a bar. The time resolution of such scintillators is sufficiently good ($\approx$5~ns) that background can be measured during gaps within the accelerator bunch structure (such as the abort gap), as well as in beam-off periods.  Since the background rate for single PE pulses from dark-current, noise, and background radiation is expected to be relatively large (about 100~Hz to 1~kHz depending on the quality of the detector elements used), we propose to add extra layers of scintillators to form coincidences with a signal in the corresponding bar of the first layer within a narrow time window ($\approx$5~ns). Muons from either pp collisions or cosmic rays could be vetoed if more than a few PE are deposited. Furthermore, by inverting this veto, these same muons could be used to align and time-in the experimental apparatus.  They would also provide a ``standard candle" against which PE depositions could be compared. 

\begin{figure}[t]
\centering
\includegraphics[width=0.4\textwidth]{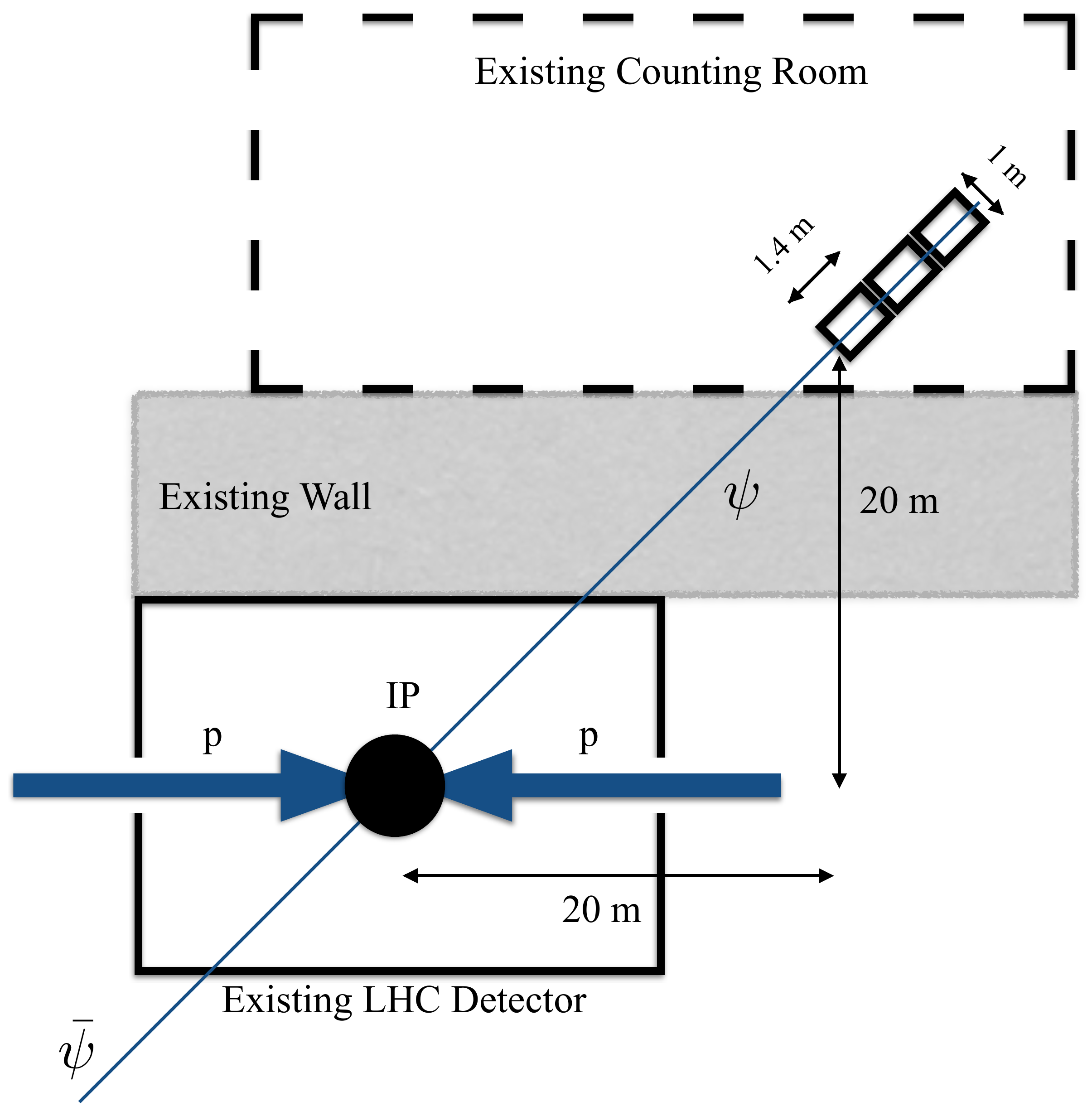}
\caption{A schematic showing the experimental setup proposed in this work. 
}
\label{fig:schematics}
\end{figure}

While the background is expected to come dominantly from the dark-current contribution, additional background from activity in the scintillator, due to background radiation and subsequent photo-multiplier afterpulsing, may also contribute significantly. These additional backgrounds can hopefully be reduced to manageable levels with additional shielding, detector optimization, and pulse-shape discrimination. This will have to be studied with small-scale detector tests {\it in situ}.  With regard to the dark-current background, we assume rates of 550~Hz, 94~Hz, 12~Hz, and 10~Hz for $N_{\rm PE}\ge 1,N_{\rm PE}\ge 2, N_{\rm PE}\ge 3, N_{\rm PE}\ge 4$, respectively in a single PMT, obtained from Fig.~65 of Ref.~\cite{pmthandbook}. Assuming an instantaneous luminosity of $2\times10^{34}$~cm$^{-2}$s$^{-1}$, and a trigger live-time of $1.5\times10^{7}$~s, we can expect $\sim 10^{10}$ background events in 300 fb$^{-1}$ for one or more PE in a single PMT.  With 200 bars needed to cover the 1 m$^2$ of area discussed above, the total background would be $\sim 10^{12}$ events in 300 fb$^{-1}$, expected to be delivered by 2022. For 3000 fb$^{-1}$, since it will be delivered with an instantaneous luminosity $1\times10^{35}$~cm$^{-2}$s$^{-1}$ the live-time will only increase by a factor of 2 and the expected background contribution would remain $\sim 10^{10}$ for one or more PE in a single PMT.  Additional discrimination can be achieved by adding two more layers of scintillators and requiring coincident PE hits. Assuming 5 ns timing resolution for the PMTs, requiring a coincidence in the second layer would reduce the background to $10^6$ coincident events with $N_{\rm PE}\ge1$ in a PMT pair of back-to-back scintillator bars. Requiring triple-incidence by adding a third layer would then bring the background to $\mathcal{O}(10)$ events with $N_{\rm PE}\ge 1$, at the cost of a moderate loss in signal efficiency.  It is possible that the slewing of small signals and/or time-of-flight differences for photons within the scintillators could degrade the timing resolution to $\sim 10$ ns, but even in this scenario the total background contribution would only increase by a factor of $\sim 4$ when triple-incidence is required.  The experimental setup with three layers is illustrated in Fig.~\ref{fig:schematics}.


In Fig.~\ref{fig:mQBounds} we show the estimated $95\%$ C.L. exclusion and $3\sigma$ sensitivity of our proposal, assuming a detector composed of three $1~{\rm m}\times 1~{\rm m}\times 1.4~{\rm m}$ layers positioned $45\degree$ away from the beam-axis. Each layer would be composed of 200 scintillator bars, and the mCP signal is one or more PE at each of the three layers within a small $\approx$5~ns window of each other. This setup could be realized if the detector is placed in either of the counting rooms at ATLAS or CMS. We estimate the signal detection efficiency by estimating the probability that a mCP signal leaves one or more PE, which we take to be Poisson distributed\footnote{The true number of PE created must be fairly large, at least $\sim$ 10, but the number we expect to observe with the tube is $N_{\rm PE}*\epsilon_{\rm eff}$, where $\epsilon_{\rm eff}$ is
the efficiency for detecting a true PE, which is about ~10\% \cite{Prinz:1998ua,pmthandbook}. Thus, the main effect on the $N_{\rm PE}*\epsilon_{\rm eff}$ distribution is from the fluctuations in how many of the true PE get detected, which is Poisson-distributed.}. The average number of PE deposited by a mCP is given by $\lambda=((Q/e)/(2\times10^{-3}))^2$ \cite{Prinz:1998ua}, assuming that the mCPs energy loss is described by the Bethe-Bloch equation \cite{Beringer:1900zz} \footnote{We checked that for  charges near $Q\approx 2\times10^{-3}$ --- when the number of interactions reaches $\mathcal{O}(1)$ ---- requiring one single hard scatter by the signal that gives 1 PE gives a sensitivity that's comparable to that given by the Bethe-Bloch equation.}. Though the Lorentz force on a mCP due to the magnetic field at either ATLAS or CMS is suppressed by $Q$, we estimate that it would produce a $\mathcal{O}(0.1-100)$~cm deviation in their trajectory over 20~m for $Q=(0.001-0.1)e$ and a momentum of $10-100$~GeV;  we have neglected this effect in the calculation of the signal acceptance. 


In this {\it Letter} we proposed a model-independent search for mCPs, which will extend sensitivity in the mass range $0.1 \lsim \MQ \lsim 100 \GeV$ by up to two orders of magnitude in electric charge over previous experiments. We estimated the potential sensitivity of this experiment to the particular realization of mCPs in ``dark QED''. The experimental setup requires a new small-scale scintillator detector nearby one of the high-luminosity interaction points at the LHC, i.e. ATLAS or CMS. Such a detector seems feasible to build for a reasonable cost with existing technology, and its placement would not interfere with existing scientific operations at the LHC.

\section*{Acknowledgments}
We thank the Abdus Salam International Center for Theoretical Physics, where this work started, for its hospitality. We also thank John Jaros, Gordan Krnjaic,  Gary Steigman and Brian Shuve for useful discussions. We thank Lance Dixon and Daniel Maitre for help with \textsc{Vrap}.  Research at Perimeter Institute is supported by the Government of Canada through Industry Canada and by the Province of Ontario through the Ministry of Research and Innovation. EI is partially supported by the Ministry of Research and Innovation - ERA (Early Research Awards) program. IY is supported in part by funds from the NSERC of Canada. 

\bibliography{milliQ_paper}

\begin{thebibliography}{27}
\expandafter\ifx\csname natexlab\endcsname\relax\def\natexlab#1{#1}\fi
\expandafter\ifx\csname bibnamefont\endcsname\relax
  \def\bibnamefont#1{#1}\fi
\expandafter\ifx\csname bibfnamefont\endcsname\relax
  \def\bibfnamefont#1{#1}\fi
\expandafter\ifx\csname citenamefont\endcsname\relax
  \def\citenamefont#1{#1}\fi
\expandafter\ifx\csname url\endcsname\relax
  \def\url#1{\texttt{#1}}\fi
\expandafter\ifx\csname urlprefix\endcsname\relax\def\urlprefix{URL }\fi
\providecommand{\bibinfo}[2]{#2}
\providecommand{\eprint}[2][]{\url{#2}}

\bibitem[{\citenamefont{Langacker and Pi}(1980)}]{Langacker:1980kd}
\bibinfo{author}{\bibfnamefont{P.}~\bibnamefont{Langacker}} \bibnamefont{and}
  \bibinfo{author}{\bibfnamefont{S.-Y.} \bibnamefont{Pi}},
  \bibinfo{journal}{Phys.Rev.Lett.} \textbf{\bibinfo{volume}{45}},
  \bibinfo{pages}{1} (\bibinfo{year}{1980}).

\bibitem[{\citenamefont{Davidson et~al.}(1991)\citenamefont{Davidson, Campbell,
  and Bailey}}]{Davidson:1991si}
\bibinfo{author}{\bibfnamefont{S.}~\bibnamefont{Davidson}},
  \bibinfo{author}{\bibfnamefont{B.}~\bibnamefont{Campbell}}, \bibnamefont{and}
  \bibinfo{author}{\bibfnamefont{D.~C.} \bibnamefont{Bailey}},
  \bibinfo{journal}{Phys.Rev.} \textbf{\bibinfo{volume}{D43}},
  \bibinfo{pages}{2314} (\bibinfo{year}{1991}).

\bibitem[{\citenamefont{Davidson et~al.}(2000)\citenamefont{Davidson,
  Hannestad, and Raffelt}}]{Davidson:2000hf}
\bibinfo{author}{\bibfnamefont{S.}~\bibnamefont{Davidson}},
  \bibinfo{author}{\bibfnamefont{S.}~\bibnamefont{Hannestad}},
  \bibnamefont{and} \bibinfo{author}{\bibfnamefont{G.}~\bibnamefont{Raffelt}},
  \bibinfo{journal}{JHEP} \textbf{\bibinfo{volume}{0005}}, \bibinfo{pages}{003}
  (\bibinfo{year}{2000}), \eprint{hep-ph/0001179}.

\bibitem[{\citenamefont{Essig et~al.}(2013)\citenamefont{Essig, Jaros, Wester,
  Adrian, Andreas et~al.}}]{Essig:2013lka}
\bibinfo{author}{\bibfnamefont{R.}~\bibnamefont{Essig}},
  \bibinfo{author}{\bibfnamefont{J.~A.} \bibnamefont{Jaros}},
  \bibinfo{author}{\bibfnamefont{W.}~\bibnamefont{Wester}},
  \bibinfo{author}{\bibfnamefont{P.~H.} \bibnamefont{Adrian}},
  \bibinfo{author}{\bibfnamefont{S.}~\bibnamefont{Andreas}},
  \bibnamefont{et~al.} (\bibinfo{year}{2013}), \eprint{1311.0029}.

\bibitem[{\citenamefont{Prinz et~al.}(1998)\citenamefont{Prinz, Baggs, Ballam,
  Ecklund, Fertig et~al.}}]{Prinz:1998ua}
\bibinfo{author}{\bibfnamefont{A.}~\bibnamefont{Prinz}},
  \bibinfo{author}{\bibfnamefont{R.}~\bibnamefont{Baggs}},
  \bibinfo{author}{\bibfnamefont{J.}~\bibnamefont{Ballam}},
  \bibinfo{author}{\bibfnamefont{S.}~\bibnamefont{Ecklund}},
  \bibinfo{author}{\bibfnamefont{C.}~\bibnamefont{Fertig}},
  \bibnamefont{et~al.}, \bibinfo{journal}{Phys.Rev.Lett.}
  \textbf{\bibinfo{volume}{81}}, \bibinfo{pages}{1175} (\bibinfo{year}{1998}),
  \eprint{hep-ex/9804008}.

\bibitem[{\citenamefont{Badertscher et~al.}(2007)\citenamefont{Badertscher,
  Crivelli, Fetscher, Gendotti, Gninenko et~al.}}]{Badertscher:2006fm}
\bibinfo{author}{\bibfnamefont{A.}~\bibnamefont{Badertscher}},
  \bibinfo{author}{\bibfnamefont{P.}~\bibnamefont{Crivelli}},
  \bibinfo{author}{\bibfnamefont{W.}~\bibnamefont{Fetscher}},
  \bibinfo{author}{\bibfnamefont{U.}~\bibnamefont{Gendotti}},
  \bibinfo{author}{\bibfnamefont{S.}~\bibnamefont{Gninenko}},
  \bibnamefont{et~al.}, \bibinfo{journal}{Phys.Rev.}
  \textbf{\bibinfo{volume}{D75}}, \bibinfo{pages}{032004}
  (\bibinfo{year}{2007}), \eprint{hep-ex/0609059}.

\bibitem[{\citenamefont{Chatrchyan et~al.}(2013)}]{CMS:2012xi}
\bibinfo{author}{\bibfnamefont{S.}~\bibnamefont{Chatrchyan}}
  \bibnamefont{et~al.} (\bibinfo{collaboration}{CMS Collaboration}),
  \bibinfo{journal}{Phys.Rev.} \textbf{\bibinfo{volume}{D87}},
  \bibinfo{pages}{092008} (\bibinfo{year}{2013}), \eprint{1210.2311}.

\bibitem[{\citenamefont{Mohapatra and Rothstein}(1990)}]{Mohapatra:1990vq}
\bibinfo{author}{\bibfnamefont{R.}~\bibnamefont{Mohapatra}} \bibnamefont{and}
  \bibinfo{author}{\bibfnamefont{I.}~\bibnamefont{Rothstein}},
  \bibinfo{journal}{Phys.Lett.} \textbf{\bibinfo{volume}{B247}},
  \bibinfo{pages}{593} (\bibinfo{year}{1990}).

\bibitem[{\citenamefont{Davidson and Peskin}(1994)}]{Davidson:1993sj}
\bibinfo{author}{\bibfnamefont{S.}~\bibnamefont{Davidson}} \bibnamefont{and}
  \bibinfo{author}{\bibfnamefont{M.~E.} \bibnamefont{Peskin}},
  \bibinfo{journal}{Phys.Rev.} \textbf{\bibinfo{volume}{D49}},
  \bibinfo{pages}{2114} (\bibinfo{year}{1994}), \eprint{hep-ph/9310288}.

\bibitem[{\citenamefont{Dubovsky et~al.}(2004)\citenamefont{Dubovsky, Gorbunov,
  and Rubtsov}}]{Dubovsky:2003yn}
\bibinfo{author}{\bibfnamefont{S.}~\bibnamefont{Dubovsky}},
  \bibinfo{author}{\bibfnamefont{D.}~\bibnamefont{Gorbunov}}, \bibnamefont{and}
  \bibinfo{author}{\bibfnamefont{G.}~\bibnamefont{Rubtsov}},
  \bibinfo{journal}{JETP Lett.} \textbf{\bibinfo{volume}{79}},
  \bibinfo{pages}{1} (\bibinfo{year}{2004}), \eprint{hep-ph/0311189}.

\bibitem[{\citenamefont{Dolgov et~al.}(2013)\citenamefont{Dolgov, Dubovsky,
  Rubtsov, and Tkachev}}]{Dolgov:2013una}
\bibinfo{author}{\bibfnamefont{A.}~\bibnamefont{Dolgov}},
  \bibinfo{author}{\bibfnamefont{S.}~\bibnamefont{Dubovsky}},
  \bibinfo{author}{\bibfnamefont{G.}~\bibnamefont{Rubtsov}}, \bibnamefont{and}
  \bibinfo{author}{\bibfnamefont{I.}~\bibnamefont{Tkachev}},
  \bibinfo{journal}{Phys.Rev.} \textbf{\bibinfo{volume}{D88}},
  \bibinfo{pages}{117701} (\bibinfo{year}{2013}), \eprint{1310.2376}.

\bibitem[{\citenamefont{Vogel and Redondo}(2014)}]{Vogel:2013raa}
\bibinfo{author}{\bibfnamefont{H.}~\bibnamefont{Vogel}} \bibnamefont{and}
  \bibinfo{author}{\bibfnamefont{J.}~\bibnamefont{Redondo}},
  \bibinfo{journal}{JCAP} \textbf{\bibinfo{volume}{1402}}, \bibinfo{pages}{029}
  (\bibinfo{year}{2014}), \eprint{1311.2600}.

\bibitem[{\citenamefont{Langacker and Steigman}(2011)}]{Langacker:2011db}
\bibinfo{author}{\bibfnamefont{P.}~\bibnamefont{Langacker}} \bibnamefont{and}
  \bibinfo{author}{\bibfnamefont{G.}~\bibnamefont{Steigman}},
  \bibinfo{journal}{Phys.Rev.} \textbf{\bibinfo{volume}{D84}},
  \bibinfo{pages}{065040} (\bibinfo{year}{2011}), \eprint{1107.3131}.

\bibitem[{\citenamefont{Holdom}(1986)}]{Holdom:1985ag}
\bibinfo{author}{\bibfnamefont{B.}~\bibnamefont{Holdom}},
  \bibinfo{journal}{Phys.Lett.} \textbf{\bibinfo{volume}{B166}},
  \bibinfo{pages}{196} (\bibinfo{year}{1986}).

\bibitem[{\citenamefont{Perl et~al.}(2009)\citenamefont{Perl, Lee, and
  Loomba}}]{Perl:2009zz}
\bibinfo{author}{\bibfnamefont{M.~L.} \bibnamefont{Perl}},
  \bibinfo{author}{\bibfnamefont{E.~R.} \bibnamefont{Lee}}, \bibnamefont{and}
  \bibinfo{author}{\bibfnamefont{D.}~\bibnamefont{Loomba}},
  \bibinfo{journal}{Ann.Rev.Nucl.Part.Sci.} \textbf{\bibinfo{volume}{59}},
  \bibinfo{pages}{47} (\bibinfo{year}{2009}).

\bibitem[{\citenamefont{Jaeckel and Ringwald}(2010)}]{Jaeckel:2010ni}
\bibinfo{author}{\bibfnamefont{J.}~\bibnamefont{Jaeckel}} \bibnamefont{and}
  \bibinfo{author}{\bibfnamefont{A.}~\bibnamefont{Ringwald}},
  \bibinfo{journal}{Ann.Rev.Nucl.Part.Sci.} \textbf{\bibinfo{volume}{60}},
  \bibinfo{pages}{405} (\bibinfo{year}{2010}), \eprint{1002.0329}.

\bibitem[{\citenamefont{Brust et~al.}(2013)\citenamefont{Brust, Kaplan, and
  Walters}}]{Brust:2013ova}
\bibinfo{author}{\bibfnamefont{C.}~\bibnamefont{Brust}},
  \bibinfo{author}{\bibfnamefont{D.~E.} \bibnamefont{Kaplan}},
  \bibnamefont{and} \bibinfo{author}{\bibfnamefont{M.~T.}
  \bibnamefont{Walters}}, \bibinfo{journal}{JHEP}
  \textbf{\bibinfo{volume}{1312}}, \bibinfo{pages}{058} (\bibinfo{year}{2013}),
  \eprint{1303.5379}.

\bibitem[{\citenamefont{Aubert et~al.}(2009)}]{Aubert:2009ae}
\bibinfo{author}{\bibfnamefont{B.}~\bibnamefont{Aubert}} \bibnamefont{et~al.}
  (\bibinfo{collaboration}{BaBar Collaboration}),
  \bibinfo{journal}{Phys.Rev.Lett.} \textbf{\bibinfo{volume}{103}},
  \bibinfo{pages}{251801} (\bibinfo{year}{2009}), \eprint{0908.2840}.

\bibitem[{\citenamefont{Aubert et~al.}(2008)}]{Aubert:2008as}
\bibinfo{author}{\bibfnamefont{B.}~\bibnamefont{Aubert}} \bibnamefont{et~al.}
  (\bibinfo{collaboration}{BaBar Collaboration}) (\bibinfo{year}{2008}),
  \eprint{0808.0017}.

\bibitem[{\citenamefont{Jaeckel et~al.}(2013)\citenamefont{Jaeckel, Jankowiak,
  and Spannowsky}}]{Jaeckel:2012yz}
\bibinfo{author}{\bibfnamefont{J.}~\bibnamefont{Jaeckel}},
  \bibinfo{author}{\bibfnamefont{M.}~\bibnamefont{Jankowiak}},
  \bibnamefont{and}
  \bibinfo{author}{\bibfnamefont{M.}~\bibnamefont{Spannowsky}},
  \bibinfo{journal}{Phys.Dark Univ.} \textbf{\bibinfo{volume}{2}},
  \bibinfo{pages}{111} (\bibinfo{year}{2013}), \eprint{1212.3620}.

\bibitem[{\citenamefont{Alwall et~al.}(2011)\citenamefont{Alwall, Herquet,
  Maltoni, Mattelaer, and Stelzer}}]{Alwall:2011uj}
\bibinfo{author}{\bibfnamefont{J.}~\bibnamefont{Alwall}},
  \bibinfo{author}{\bibfnamefont{M.}~\bibnamefont{Herquet}},
  \bibinfo{author}{\bibfnamefont{F.}~\bibnamefont{Maltoni}},
  \bibinfo{author}{\bibfnamefont{O.}~\bibnamefont{Mattelaer}},
  \bibnamefont{and} \bibinfo{author}{\bibfnamefont{T.}~\bibnamefont{Stelzer}},
  \bibinfo{journal}{JHEP} \textbf{\bibinfo{volume}{1106}}, \bibinfo{pages}{128}
  (\bibinfo{year}{2011}), \eprint{1106.0522}.

\bibitem[{\citenamefont{Ball et~al.}(2013)\citenamefont{Ball, Bertone,
  Carrazza, Deans, Del~Debbio et~al.}}]{Ball:2012cx}
\bibinfo{author}{\bibfnamefont{R.~D.} \bibnamefont{Ball}},
  \bibinfo{author}{\bibfnamefont{V.}~\bibnamefont{Bertone}},
  \bibinfo{author}{\bibfnamefont{S.}~\bibnamefont{Carrazza}},
  \bibinfo{author}{\bibfnamefont{C.~S.} \bibnamefont{Deans}},
  \bibinfo{author}{\bibfnamefont{L.}~\bibnamefont{Del~Debbio}},
  \bibnamefont{et~al.}, \bibinfo{journal}{Nucl.Phys.}
  \textbf{\bibinfo{volume}{B867}}, \bibinfo{pages}{244} (\bibinfo{year}{2013}),
  \eprint{1207.1303}.

\bibitem[{\citenamefont{Anastasiou et~al.}(2004)\citenamefont{Anastasiou,
  Dixon, Melnikov, and Petriello}}]{vrap}
\bibinfo{author}{\bibfnamefont{C.}~\bibnamefont{Anastasiou}},
  \bibinfo{author}{\bibfnamefont{L.~J.} \bibnamefont{Dixon}},
  \bibinfo{author}{\bibfnamefont{K.}~\bibnamefont{Melnikov}}, \bibnamefont{and}
  \bibinfo{author}{\bibfnamefont{F.}~\bibnamefont{Petriello}},
  \bibinfo{journal}{Phys.Rev.} \textbf{\bibinfo{volume}{D69}},
  \bibinfo{pages}{094008} (\bibinfo{year}{2004}), \eprint{hep-ph/0312266},
  \urlprefix\url{http://www.slac.stanford.edu/~lance/Vrap/}.

\bibitem[{\citenamefont{Artoisenet et~al.}(2008)\citenamefont{Artoisenet,
  Maltoni, and Stelzer}}]{Artoisenet:2007qm}
\bibinfo{author}{\bibfnamefont{P.}~\bibnamefont{Artoisenet}},
  \bibinfo{author}{\bibfnamefont{F.}~\bibnamefont{Maltoni}}, \bibnamefont{and}
  \bibinfo{author}{\bibfnamefont{T.}~\bibnamefont{Stelzer}},
  \bibinfo{journal}{JHEP} \textbf{\bibinfo{volume}{0802}}, \bibinfo{pages}{102}
  (\bibinfo{year}{2008}), \eprint{0712.2770}.

\bibitem[{\citenamefont{Dawson and Hedberg}(2004)}]{Dawson:2004pta}
\bibinfo{author}{\bibfnamefont{I.}~\bibnamefont{Dawson}} \bibnamefont{and}
  \bibinfo{author}{\bibfnamefont{V.}~\bibnamefont{Hedberg}}
  (\bibinfo{year}{2004}), \eprint{ATL-TECH-2004-001}.

\bibitem[{\citenamefont{Beringer et~al.}(2012)}]{Beringer:1900zz}
\bibinfo{author}{\bibfnamefont{J.}~\bibnamefont{Beringer}} \bibnamefont{et~al.}
  (\bibinfo{collaboration}{Particle Data Group}), \bibinfo{journal}{Phys.Rev.}
  \textbf{\bibinfo{volume}{D86}}, \bibinfo{pages}{010001}
  (\bibinfo{year}{2012}).

\bibitem[{\citenamefont{Burle Industries Inc.~(Lancaster}(1980)}]{pmthandbook}
\bibinfo{author}{\bibfnamefont{P.}~\bibnamefont{Burle Industries
  Inc.~(Lancaster}}, \emph{\bibinfo{title}{Photomultiplier Handbook: Theory,
  Design, Application}} (\bibinfo{publisher}{Burle Industries},
  \bibinfo{year}{1980}),
  \urlprefix\url{http://psec.uchicago.edu/links/Photomultiplier_Handbook.pdf}.

\end{thebibliography}
\end{document}